\documentclass{optica-article}

\journal{optcon}

\usepackage{lineno}

\newcommand{\pmSqHz}{\ensuremath{{\mathrm{pm}}/{\sqrt{\mathrm{Hz}}}}}

\begin{document}

\title{Sub-nanometer measurement of transient structural changes in dye-doped polystyrene microspheres }

\author{Pegah Asgari,\authormark{1} Itır Bakış Dogru Y\"{u}ksel,\authormark{1} Gerhard A. Blab,\authormark{1} Hans C. Gerritsen,\authormark{1} and Allard P. Mosk\authormark{1,*}}

\address{\authormark{1} Debye Institute for Nanomaterials Science, Utrecht University, The Netherlands\\
}

\email{\authormark{*}a.p.mosk@uu.nl} 



\begin{abstract*}
We  present an interferometric spectral-domain optical coherence tomography microscopy setup to detect structural changes using interference of light reflected from different interfaces of the sample. We induce a reproducible nanometer-scale size change in dye-doped 10-µm polystyrene microspheres by the release of Stokes shift energy of dye molecules inside the microspheres, excited by a modulated 532-nm laser. The resulting optical path length difference was measured with a sensitivity of $0.4~\pmSqHz$ limited by photodetection noise, and reveals elastic as well as inelastic responses, which opens up possibilities for measuring the response of cell-sized biological objects.
\end{abstract*}

\section{Introduction}
The microscopic physical properties of single cells attract significant research interest, because of the ability to measure e.g.  stiffness and motility of cells \cite{kwon2020comparison,iida2017cell}. The methods to detect these physical properties are usually invasive, such as AFM to measure the stiffness \cite{iida2017cell,viji2019nano} or electrode and fluorescence based \cite{kralj2012optical} electrophysiology. Label-free optical techniques could provide an alternative to non-invasively recording of neuronal activities.

Quantitative measurement of the optical phase is emerging as a powerful, label-free approach to imaging cells and tissues, especially as it combines qualities found in microscopy, holography, and light scattering techniques: nanoscale sensitivity of morphology and dynamics, non-destructive imaging of completely transparent structures and quantitative signals based on intrinsic contrast \cite{park2018quantitative}. 
By combining holography and microscopy one can perform sensitive measurements of the thickness and refractive index of biological specimens, with sub-wavelength accuracy \cite{hariharan2003optical}. Early on, such quantitative phase measurements were performed by single point scanning techniques, building on the advancement of optical coherence tomography (OCT) \cite{huang1991optical}. Two classes of OCT are swept-source OCT (SS-OCT), which uses a tunable swept laser \cite{wang2020choroidal,yu2018cataract,zheng2019age} and spectral-domain OCT (SD-OCT) which uses a broadband light source \cite{nguyen2011spectral,van2020deep}. While SS-OCT has some advantages in imaging applications, SD-OCT is preferable in applications with time-dependent signals as it can take an entire spectrum in a single shot. 
In 2001, Yang et al. developed an OCT system that could detect motion within a cell with ~3.8~nm sensitivity for a slow sampling \cite{yang2001phase}. Many studies were carried out using the same methods to detect the optical and mechanical change of neurons during action potentials (APs) \cite{akkin2007depth,batabyal2017label}. In all the studies conducted to detect the cellular activities and motion, the reference beam was a mirror in a separate beam path such as a Michelson configuration or common path configurations, for instance, one side of the coverslip, which holds the sample. In these configurations besides the influence of the noise on the sensitivity of the experiment, the motion of the object relative to the coverslip exhibit crosstalk with changes within the object itself. This crosstalk problem becomes worse if the object is closer to the coverslip than the axial resolution of the SD-OCT, as in this case it is not possible to separate the desired signal from the whole object's motion.

In this work, we aim to measure sub-nanometer transient path length changes of cell-sized objects, where we use polymer microspheres as a model system. We have constructed an OCT setup without an external reference beam, instead employing a common path configuration in which the phase is retrieved from the interference of the reflected light from the top and bottom surface of the microsphere. To show the high sensitivity of SD-OCT in a dynamical setting we use a high sampling rate of 80~kHz. 

As reference objects of cellular size which can be deformed in a controllable way in real-time, we use polystyrene (PS) microspheres, with a diameter of 10 $\mu$m, doped with fluorescent dye. When the dye molecules absorb light, part of the energy (corresponding to the Stokes shift) is released as heat \cite{kasap2013optoelectronics}. The OCT system does not detect fluorescence, as the fluorescent light is not coherent with the source \cite{ntziachristos2010going}, moreover, the fluorescent light is in a different wavelength range. The released heat changes the density, volume, and refractive index of the object and we measure the resulting optical path length change. 

\section{Methods and Materials}
\subsection{Spectral-domain OCT setup}
Light from a superluminescent diode (Superlum Inc., $\lambda_0$=840~nm, $\Delta\lambda$=160~nm) is divided by a fiber coupler, where one output is coupled into an inverted microscope (Nikon E2000 with a 20x NA=0.75 objective), which contains the sample (Fig.~1). The reflected light from the sample is coupled back into the same fiber, and part of it passes the fiber coupler to reach a spectrometer (Wasatch Photonics, Cobra S-800). The spectrum is recorded by a line camera (E2V OCTAPLUS) with an 80-kHz line rate and a spectral resolution of 0.15~nm. An 80/20 beamsplitter inside the microscope side port allows for simultaneous sample visualization and interferometry. To expand the microspheres by heat, a fiber-coupled, 532-nm laser (Coherent Obis) is connected to the excitation port of the microscope, overlapping with the OCT beam using a Nikon G-2B cube (Long-pass Emission). This cube normally has a dichroic mirror, an excitation filter, and a transmission filter, however in this experiment we have removed the transmission filter, so the whole bandwidth of the SLD is transmitted through the cube. We modulated the excitation laser with a function generator, controlled by a microcontroller (Arduino), with a modulation period of 88~ms and a duty cycle of 50\%. The excitation light is collimated and focused through the (underfilled) microscope objective, so that the beam waist on the sample is $\approx$4 $\mu$m and the Rayleigh length is larger than 10$\mu$m.

\begin{figure}[ht!]
\centering\includegraphics[width=11cm]{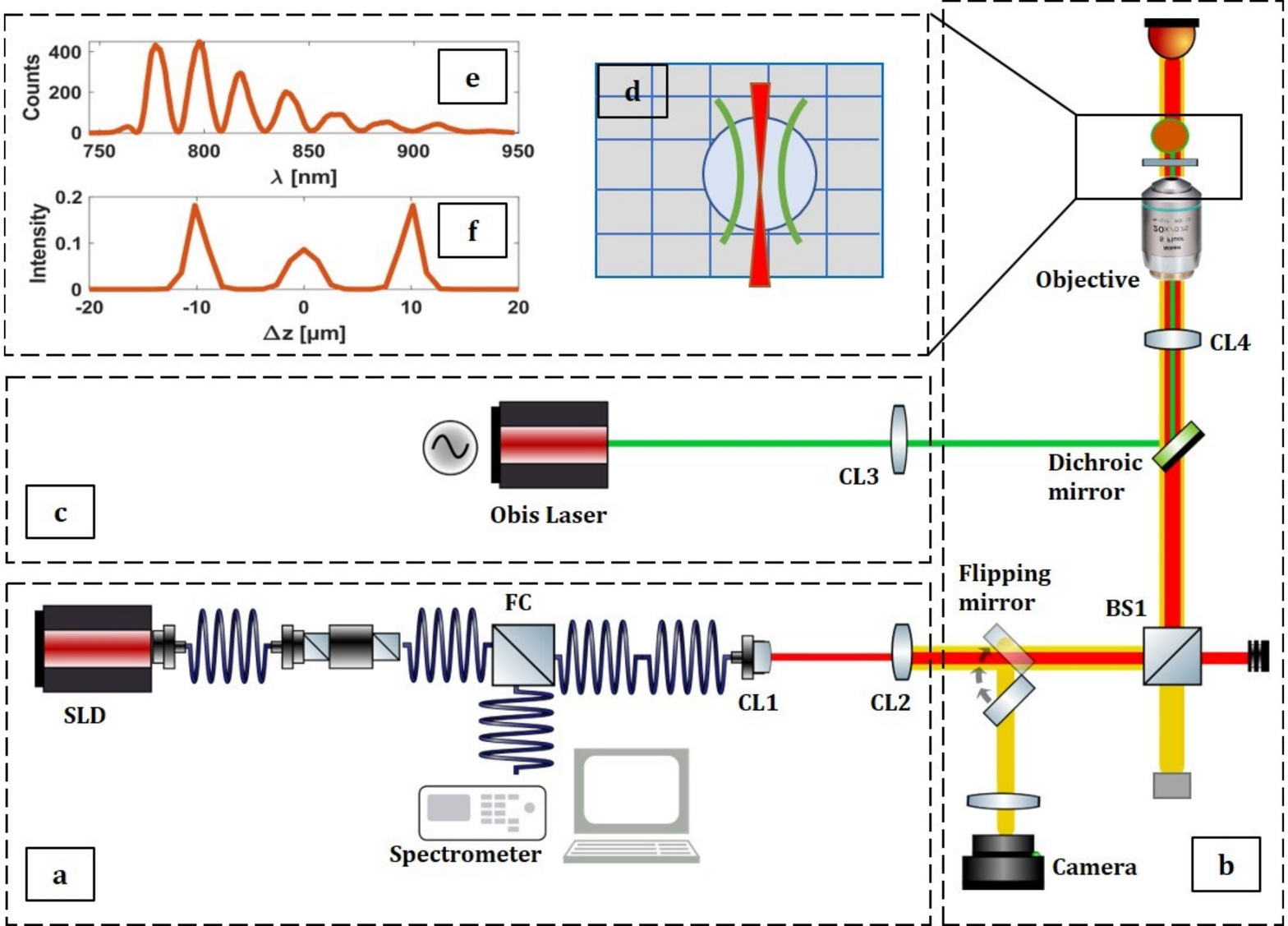}
\caption{Setup of the SD-OCT system. (a) The light source and detection part that includes a superluminescent light source (SLD), a fiber coupler (FC) (two practical options 90:10 or 50:50), a collimator (CL1), and spectrometer (SP). (b) Microscopy section where the light is directed to the sample by a collimated lens (CL2), BS1 (P), microscope objective, and arrives at the microsphere.  (c) excitation arm with a 532-nm laser, with collimation lens (CL3). (d) Schematic of the sample (e) Measured interferogram formed by the top and bottom reflections from the microsphere and (f) its intensity Fourier transform}
\end{figure}

The sample is a single polystyrene microsphere (FluoSpheres, ThermoFisher) doped with orange fluorescent dye. The excitation peak is at 540~nm, with the emission peak at 560~nm. The microsphere was suspended in 1\% agarose gel which keeps the microspheres fixed in space and reduces the light-pressure induced motion of the microspheres during heating. 
The desired contribution to our signal is formed by the reflections from the bottom and top surface of the microsphere and their interference term (Fig.1). The power spectral density due to two reflectors, here arbitrarily identified as reference and sample, is given by \cite{choma2005spectral,goodman2015statistical},

\begin{equation}
I(k)= S(k)\sqrt{R_r R_s}\cos(2nk [z_0 +\delta z (t)])
\end{equation}

Where $z_0 + \delta z$ is the distance between the reference and sample reflectors, k is wave-number and n is the refractive index between the reference and sample reflectors. $z_0$ denotes the initial position of the sample reflector, which is determined within the source coherence length, while $\delta z(t)$ accounts for time-dependent optical path length deviations. $S(k)$ is the source spectral power density function; $R_r$ and $R_s$ are the reference and sample reflectivity, respectively. Fourier transformation of $I(k)$ yields an A-scan $\mathfrak{I}(nz)$ that has peak values at $2nz_0$,

\begin{equation}
\mathfrak{I}(\pm2n z_0)= E(2n z_0) \sqrt{R_r R_s} \exp{[2 k_c n \delta z(t)]},
\end{equation}

Where $E(2n z_0)$ is the unity-amplitude coherence envelope function and $k_c$  is the source center wave-number. Since the phase of $\mathfrak{I}(\pm2n z_0)$ is a linear function of $\delta z$, time-varying changes in the distance between the sample and reference reflectors can be measured with reference to the distance at time $t_0$,

\begin{equation}
\delta \varphi(t)=\angle \mathfrak{I}(\pm2n z_0,t) - \angle \mathfrak{I}(\pm2n z_0,t_0)=2k_c n \delta z,
\end{equation}

Eq. 3 shows the time varying phase of the interferograms where $\angle$ denotes angle operator. 

\section{Results}

\subsection{Transient structure analysis of microspheres}
In Fig. 2 we show the measured changes in OPL while illuminating a single microsphere with the modulated excitation laser.  We observe that every pulse of the excitation laser induces a temporary reduction of the OPL, even though a thermal expansion of the microsphere is expected.  The change of the OPL is proportional to the excitation laser power. To explain the negative sign of the change we invoke the Clausius-Mossotti equation which relates the refractive index to density of the object, leading to a decrease of the refractive index with temperature \cite{sasabe1972effects}.

\begin{figure}[ht!]
\centering\includegraphics[width=11cm]{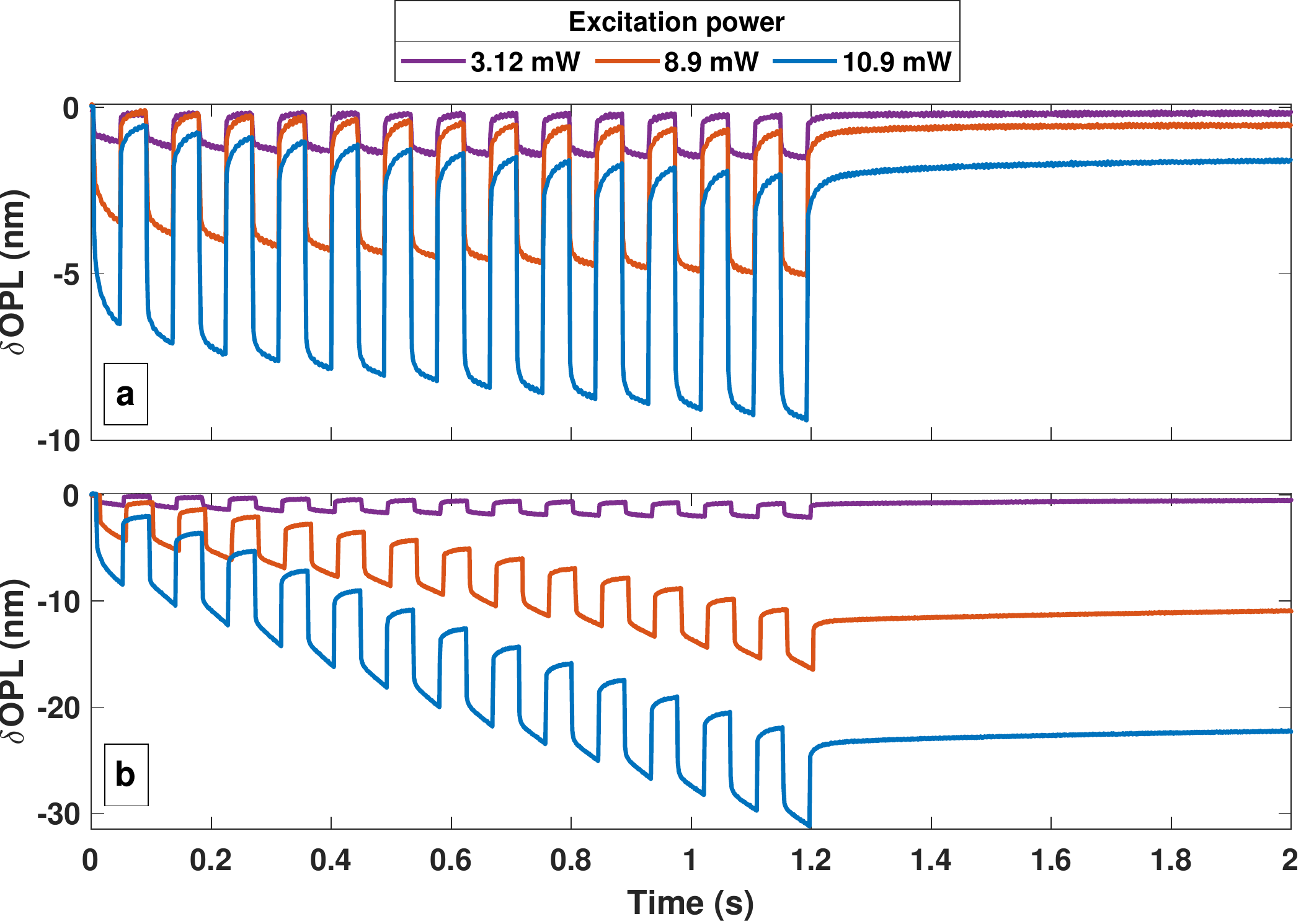}
\caption{Optical path length of two microspheres exposed to various power, as example of two different microspheres behavior (a) Microsphere showing a low amount of inelastic creep. (b) Microsphere showing significant creep}
\end{figure}

Since the OPL is the product of the geometric length of the optical and the refractive index, therefore the change in OPL as function of temperature T is,

\begin{equation}
\delta \mathrm{OPL}=2n\delta z(T) + 2z \delta n(T).
\end{equation}

Using the Clausius-Mossotti relation we find,

\begin{equation}
\frac{d \delta \mathrm{OPL}}{dT}=2 [- \frac{(n^2-1)(n^2+2)}{2n}+n]\frac{d\delta z}{dT}
\end{equation}
where the first term in brackets represents the effect of change in the refractive index and the second one the thermal expansion of the radius of the microsphere on the OPL. In the case of a polystyrene sample with n=$1.57$,  the term in brackets evaluates to $-0.53$, as the effect of change  in refractive index is dominant.

Besides the step changes, we observe in Fig. 2 slow change in the baseline of OPL, which is uniformly decreasing for all particles observed, although in different amounts. This slow change is not due to a drift in position, as this would cause a loss of fringe visibility which we do not observe. We tentatively attribute the slow drift to inelastic deformation \cite{brinson2008polymer,riande2000flow} of the sphere which is heated inhomogeneously by our laser. This slow creep deformation depends on the molecular properties of the polymer and the stress/strain history that polymer has been through \cite{brinson2008polymer,lin2011polymer}. 
In Fig.~2 we show the response of two microspheres from the same batch. Some microspheres show predominantly elastic behavior as shown in Fig.~2a, while in Fig. 2b we show an example of a microsphere that undergoes significant amount of permanent path length change. This strong difference in the inelastic behavior could be attributed to a possible different annealing history of the two microspheres. The amplitude of the fast elastic response is equal in all observed of microspheres. 

To rule out contributions of absorption elsewhere, we repeated the experiments non-doped PS microspheres at the same size and kept all the parameters constant. From the response curves in Fig.~3 (which are representative of several different microspheres) we see that the dye-doped microspheres show transient changes that follow the excitation laser profile, while samples without dye showed neither ageing nor any elastic response. 

\begin{figure}[ht!]
\centering\includegraphics[width=11cm]{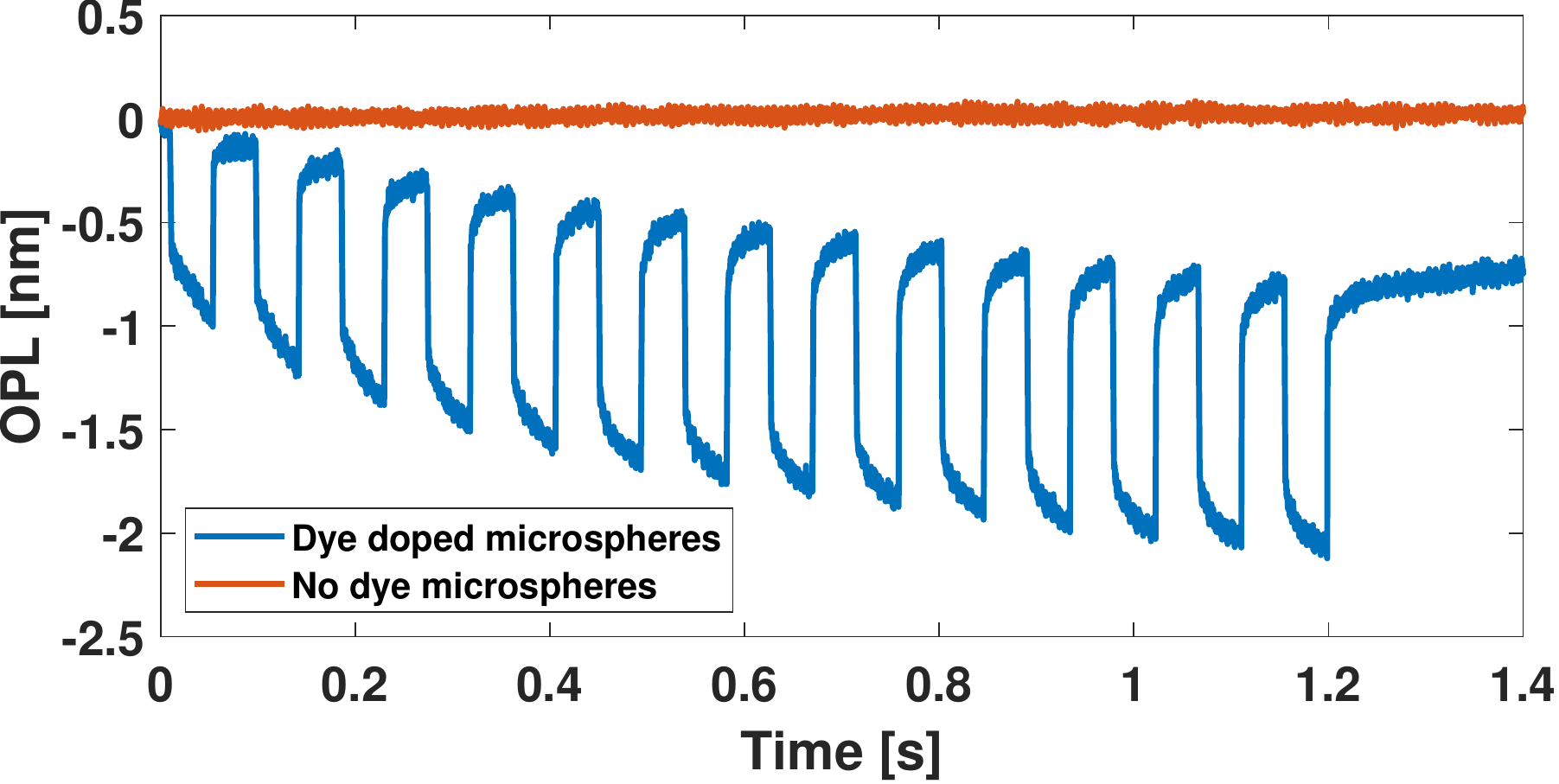}
\caption{Comparison of OPL change in dye-doped and non-absorbing microspheres experiencing the same laser power of 3.12~mW.}

\end{figure}

\subsection{Sensitivity}
We define sensitivity as the change in the $\delta \mathrm{OPL}$ that corresponds to a signal to noise ratio (SNR) of 1. The SD-OCT setup samples the spectra at 80~kHz. To characterize the sensitivity, we show the measured amplitude spectral density of the OPL in the absence of laser excitation (Fig. 4). The spectral sensitivity averaged over the relevant interval  of 0.5-40~kHz is $0.4~\pmSqHz$. As a result, at a 80-kHz sampling rate and with 2.7~mW IR power on the sample, the RMS noise of the OPL is 25~pm, while with a reduced power of 0.7~mW  we reach 40~pm.

\begin{figure}[ht!]
\centering\includegraphics[width=11cm]{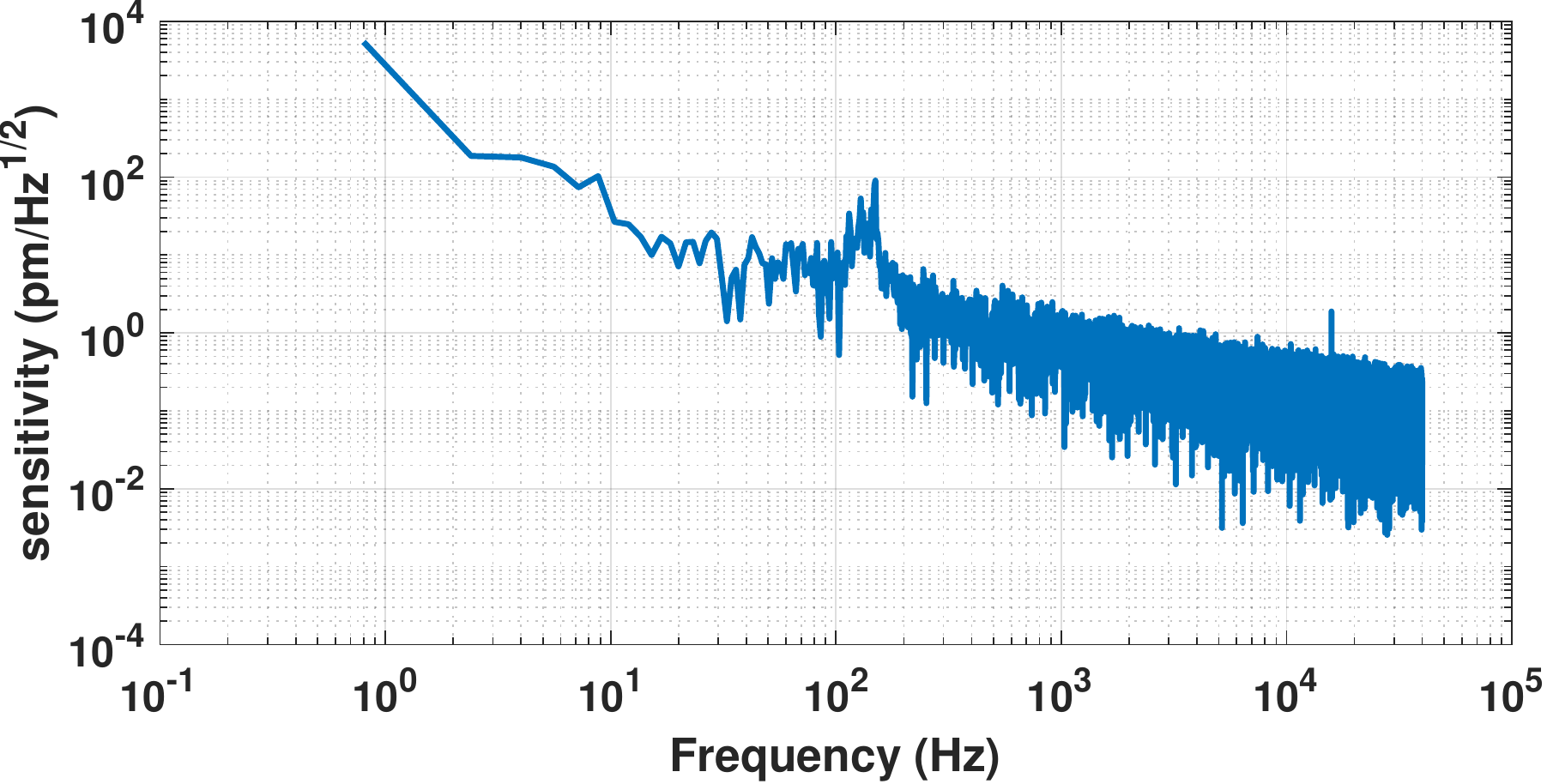}
\caption{Measured noise spectrum of our SD-OCT setup, obtained by Fourier transforming the measured OPL of a non-doped sphere, without excitation light.}
\end{figure}

\begin{figure}[ht!]
\centering\includegraphics[width=11cm]{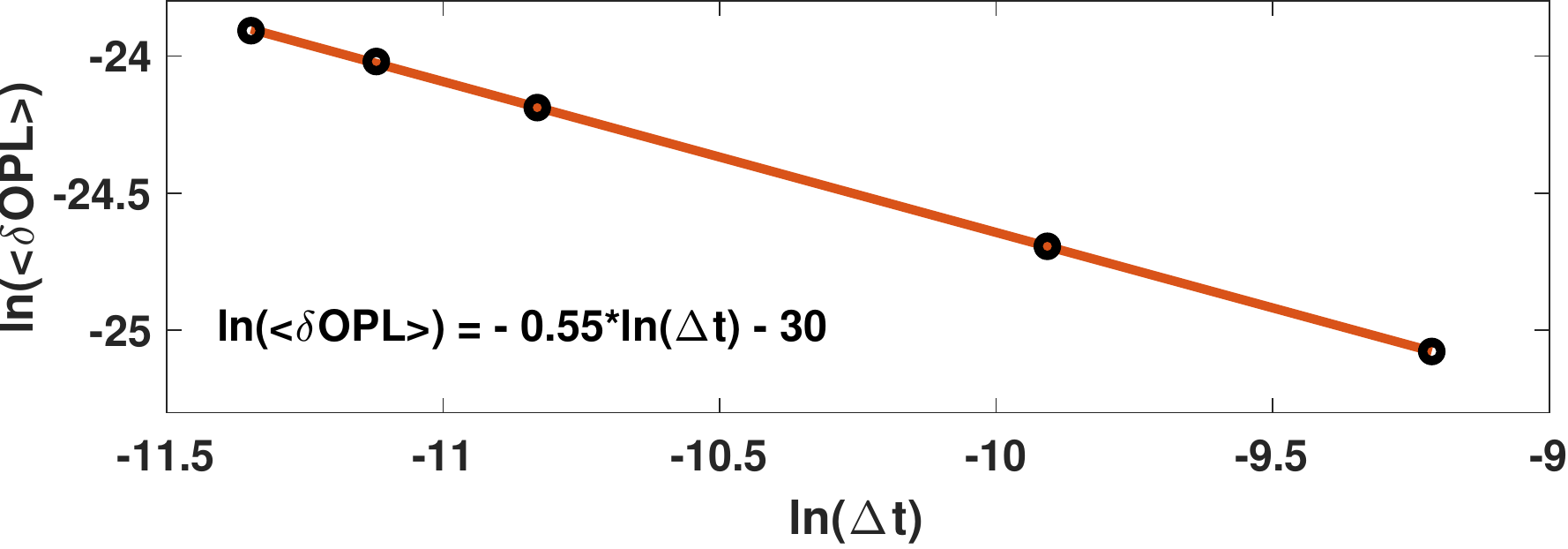}
\caption{Semilogarithmic plot of sensitivity of the $\delta \mathrm{OPL}$ versus integration time $\Delta t$. The   power of SLD on the sample is 0.7~mW.}
\label{fig:sensitivity}
\end{figure}

To verify photodetection-noise limited operation we show the sensitivity as a function of integration time in a semilogarithmic plot in Fig. \ref{fig:sensitivity} which has the a slope of $-0.55$ close to the value of $-0.5$ characteristic of shot noise \cite{choma2005spectral}. 

\section{Conclusion and discussion}
We demonstrate a sensitive SD-OCT to measure sub-nanometer transient configurational alterations. We use a superluminescent light source with broad spectral bandwidth combined with a high-speed and high-resolution spectrometer. The system can detect changes in diameter of PS microspheres with a photodetection-noise limited sensitivity of $0.4~\pmSqHz$, which reveals the effects of thermal expansion and annealing in dye-doped microspheres caused by mild laser heating. This experiment paves the road to detecting sub-nm dimensional changes in cells and similarly sized objects, decoupled from their center of mass motion.


\section*{Acknowledgments}
The authors would like to thank Dave van den Heuvel, Aron Opheij, Dante Killian, and Paul Jurrius for technical support, and Arnout Imhof, Dries van Oosten and Tayebeh Saghaei for discussions. This research is supported by the Dutch Research Council (NWO) (Vici 68047618 and the Neurophotonics program 6NEPH03).



\bibliography{biblio}






\end{document}